# Gate-tuned normal and superconducting transport at the surface of a topological insulator

Benjamin Sacépé[1,2§*], Jeroen B. Oostinga[1,2*], Jian Li[3], Alberto Ubaldini[1], Nuno J.G. Couto[1,2], Enrico Giannini[1] & Alberto F. Morpurgo[1,2#]

[1]*Département de Physique de la Matière Condensée, University of Geneva, 24 Quai Ernest-Ansermet, 1211 Genève 4, Switzerland*

[2]*Group of Applied Physics, University of Geneva, 24 Quai Ernest-Ansermet, 1211 Genève 4, Switzerland*

[3]*Département de Physique Théorique, University of Geneva, 24 Quai Ernest-Ansermet, 1211 Genève 4, Switzerland*

[§]*Present address : Institut Néel, CNRS and Université Joseph Fourier, BP 166, 38042 Grenoble, France*

[*]*These authors contributed equally to this work*

[#]*Email: Alberto.Morpurgo@unige.ch*

**Three-dimensional topological insulators are characterized by the presence of a bandgap in their bulk and gapless Dirac fermions at their surfaces. New physical phenomena originating from the presence of the Dirac fermions are predicted to occur, and to be experimentally accessible via transport measurements in suitably designed electronic devices. Here we study transport through superconducting junctions fabricated on thin $Bi_2Se_3$ single crystals, equipped with a gate electrode. In the presence of perpendicular magnetic field B, sweeping the gate voltage enables us to observe the filling of the Dirac fermion Landau levels, whose character evolves continuously from electron- to hole-like. When B=0, a**

**supercurrent appears, whose magnitude can be gate tuned, and is minimum at the charge neutrality point determined from the Landau level filling. Our results demonstrate how gated nano-electronic devices give control over normal and superconducting transport of Dirac fermions at an individual surface of a three-dimensional topological insulator.**

## Introduction

The classification of crystalline solids in terms of their electronic properties as metals and insulators is one of the early successes of quantum mechanics. Metals are materials where a finite density of states at the Fermi energy enables electrons to move freely, and to conduct electricity. On the contrary, when the Fermi energy lies in an energy gap with no electronic state available, electrons cannot propagate, causing the insulating behaviour. Remarkably, it has been recently discovered[1,2] that this long established classification is incomplete: topological insulators (TIs), realized in compounds with very strong spin-orbit interaction, represent a third class of materials, possessing an energy gap in their bulk –similarly to insulators- and electronic states that remain ungapped –like in a metal- at their surfaces.

The nature of the surface states in TIs depends on their dimensionality. In two-dimensional (2D) TIs, they consist of 1D helical modes propagating at the system edges[1,2], whose existence has been revealed in transport experiments[3,4]. In the 3D case, the surface states are 2D gases of Dirac fermions[5-9]. Following theoretical predictions[6,9], Dirac fermions have been observed in several bismuth-based materials in angle-resolved photo-emission experiments[10-12] and scanning tunnelling spectroscopy[13,14]. Contrary to the 2D case, however, for 3D TIs probing and controlling Dirac fermions in transport measurements[15-19] is difficult, because a large parallel conductance originating from bulk states is usually present[18,20-23]. This hinders the use of nano-fabricated devices based on 3D TIs, which is essential to gain control over the Dirac fermions and to probe some of their most unique properties.

Here we study transport through thin (~10 nm) $Bi_2Se_3$ layers exfoliated from single-crystals and transferred onto (doped) $Si/SiO_2$ substrates acting as gates, and we show that the ability to modulate the carrier density is a powerful tool to investigate Dirac surface fermions. Since superconductor/TI junctions are of interest for the investigation of so-called Majorana fermions[24,25], the $Bi_2Se_3$ layers were contacted with two closely separated Al/Ti superconducting electrodes ($T_c$~1 K; see Fig. 1), enabling us to investigate in a same device the normal state transport at high magnetic field $B$, and the possibility to gate-control induced superconductivity at $B = 0$.

## Results

**Gate-tuned Shubnikov-de Haas oscillations of Dirac fermions.** We realized several similar devices and here we discuss data measured on one of them that are representative of the overall behaviour observed (data from the other devices are shown in the Supplementary Note 1). We first discuss the dependence of the normal state resistance measured at $T = 4.2$ K on gate voltage $V_g$ and magnetic field $B$, focusing on the qualitative aspects of the data, which are sufficient to conclude that Dirac surface fermions are present and can be gate controlled. The sample resistance is $R$=70 Ω and shows a 15-20% variation when $V_g$ is swept from -80 V to +50 V, roughly symmetrical with respect to $V_g^{CN}$ ~ -10 V (see Fig. 2a). This non-monotonic dependence is a manifestation of ambipolar transport involving both electrons and holes, and indicates that the Fermi level at the $Bi_2Se_3$ surface is located in the gap between valence and conduction bulk bands[26,27].

With increasing *B*, the resistance also increases, reaching values close to 1 kΩ at 15 T (Fig. 2b). Oscillations periodic in *1/B* are clearly visible (Fig. 2c), as it is characteristic for the Shubnikov-de Haas (SdH) effect[28] (Fig.2b). The oscillations are gate-voltage dependent and disperse in opposite directions when $V_g$ is swept across $V_g^{CN}$, which indicates that the character of the charge carriers changes from electron ($V_g > V_g^{CN}$) to hole ($V_g < V_g^{CN}$). These oscillations are not present when the magnetic field is applied parallel to the surface of the $Bi_2Se_3$ flake (see Supplementary Note 2). Additional information can be obtained if the amplitude of the oscillations is enhanced by looking at the quantities $-\frac{d^2R}{dB^2}(B,V_g)$ and $-\frac{d^2R}{dV_g^2}(B,V_g)$ (Fig. 3a and Fig. 3b respectively) or by subtracting the positive magnetoresistance background (see Supplementary Note 3). The two methods give fully consistent results, and here we focus on the discussion of the second derivative which is free from any possible arbitrariness associated to the background subtraction. Fig. 3a shows that –next to the oscillations dispersing with $V_g$ – features that do not depend on gate voltage are also present. These features are also periodic in *1/B*: indeed the Fourier spectrum of the data plotted versus *1/B* (Fig. 3c) shows a peak dispersing with gate voltage as well as a peak whose position is $V_g$-independent. Interestingly, at $V_g = 0$ V the frequency of the two peaks –the one dispersing with gate voltage and the $V_g$-independent one- are nearly the same.

These observations give clear indications as to the nature of the electronic states that are involved. Finding an electron-to-hole crossover that occurs in a small gate voltage range (~ 10 V) –which correspond to a rather small amount of accumulated charge (few times $10^{11}$ cm$^{-2}$)- implies that the electronic states responsible for the SdH oscillations cannot be in the bulk band of the material. Since the bulk bands are separated by a gap of ~300 meV [12], crossing over from electron to holes would require a change in gate voltage that is well more than one order of magnitude larger. A continuous crossover from electron to hole SdH is known to occur (and has been observed) only in gapless systems of chiral fermions (e.g., graphene or bilayer graphene[29]). Indeed gapless Dirac fermions are known to be present at the surface of $Bi_2Se_3$ [9,12,13,14]. The electron-to-hole crossover as a function of gate voltage that we observe in the SdH oscillations, therefore, is a clear indication of the presence of Dirac fermions in our devices. It also confirms, as it was already concluded from the observation of ambipolar transport at *B=0* T, that the Fermi level at the surface is inside the gap of the bulk bands of $Bi_2Se_3$.

The conclusion that Dirac fermions are responsible for the $V_g$-dependent SdH oscillations also explains the presence of SdH oscillations with a $V_g$-independent frequency. In fact, Dirac fermions are not only present on the bottom surface of the $Bi_2Se_3$ flake, but also on the top surface. Since this top surface is further away from the gate electrode, it is electrostatically screened by carriers that are accumulated on the bottom surface and by states that are present in the "bulk" of the flake (see below). Accordingly, sweeping the gate voltage does not lead to a change in the carrier density on the top surface, and the frequency of the SdH oscillations associated to these carriers is constant. The finding that at $V_g$=0 V, the $V_g$-independent and the $V_g$-dispersing oscillations have roughly the same frequency can be explained naturally, since when no gate voltage is applied the two surfaces of the $Bi_2Se_3$ flake host approximately the same carrier density. This finding could not be explained if the two families of SdH oscillations had different origin. We conclude that the presence of surface Dirac fermions at the surface of $Bi_2Se_3$ explains all the qualitative aspects of our observations –continuous evolution from electron to hole SdH, the presence of two families of SdH with different response to the gate voltage and having the same frequency at $V_g$=0 V, and the observation of ambipolar transport. No other known scenario can account for these findings.

**Bulk contribution to the magnetoresistance.** Having established the presence of Dirac surface fermions from qualitative aspects of the data only is important, because it ensures that this conclusion does not depend on any specific assumption that needs to be made to interpret detailed aspects of the data quantitatively. To proceed with such a more quantitative analysis, we first note that in our devices another transport channel in parallel to the surface states is present, which gives a dominant contribution to the measured conductance. Indeed, the device resistance –approximately 70 Ω- is much smaller than what can be expected from the presence of surface Dirac fermions alone. To estimate the conductivity of Dirac fermions, a carrier mobility $\mu$ of 1000-2000 cm$^2$/Vs can be taken, corresponding to the best values reported in the literature for Bi$_2$Se$_3$ [26,27] (see also below). We then find that, at high gate voltage, the parallel conductivity is almost 10 times larger than the surface conductivity (see Supplementary Note 4). The surface Dirac states therefore coexist with other states in the system present inside the Bi$_2$Se$_3$ flake, at the same energy (i.e., inside the gap between bulk conduction and valence band). We attribute the origin of these states to the presence of an impurity band[21] caused by the disorder present in Bi$_2$Se$_3$, which is known to be large even in the best crystals. Indeed, a rather large density of states inside the gap, coexisting with the surface Dirac fermion states, is reproducibly seen in scanning tunnelling spectroscopy experiments[14,30].

The density of states $\nu_{Bulk}$ coexisting with the Dirac surface states can be estimated by comparing the density of charge extracted from the SdH frequency $B_F$, to that obtained from the known capacitance to the gate voltage. At $V_g$ - $V_g^{CN}$ = -70 V, for instance, the frequency of the SdH oscillations corresponds to a density $n= B_F e/h$ = 1.05 10$^{16}$ m$^{-2}$ while $C(V_g - V_g^{CN})$ = 5.1 10$^{16}$ m$^{-2}$. The discrepancy is only apparent, because with changing $V_g$ charge is accumulated near the surface on a depth given by the Thomas-Fermi screening length ($\lambda_{TF}$ ~ 3-4 nm) and the occupation of states in the impurity band reduces the density of Dirac fermions $n_{Dirac}(V_g)$ to (only a fraction of the accumulated charge occupies Dirac fermion states):

$$n_{Dirac}(V_g) = \pi(\hbar v_F)^2 \left[ -\nu_{Bulk}\lambda_{TF} + \sqrt{(\nu_{Bulk}\lambda_{TF})^2 + \frac{n_{Tot}}{\pi(\hbar v_F)^2}} \right]^2 \quad (1)$$

with $n_{Tot} = C|V_g - V_g^{CN}|/e$ and $v_F = 5 \; 10^5$ m/s (see Supplementary Note 5). By imposing that at $V_g$ = -80 V $n_{Dirac}(V_g)$ coincides with the density extracted from the fan diagram, we obtain $\nu_{Bulk}\lambda_{TF}$=3.6 $10^{13}$ cm$^{-2}$eV$^{-1}$, corresponding to $\nu_{Bulk}$~7 $10^{19}$ cm$^{-3}$eV$^{-1}$, comparable to literature estimates for similar crystals[21]. We can then use Eq. (1) to calculate the frequency of the SdH oscillations for all values of $V_g$, $B_F(V_g) = \frac{h}{e}n_{Dirac}(V_g)$, and we find quantitative agreement with the position of the peak in the Fourier spectrum with no adjustable parameters (see the blue line in Fig. 3c). Eq. (1) also naturally reproduces the $V_g$ dependence of the conductivity ($\delta\sigma(V_g)$) at $B$=0. The inset of Fig. 2a shows that $\delta\sigma(V_g) = n_{Dirac}(V_g)e\mu$ is in excellent agreement with the measured data with $\mu$ = 2500 cm$^2$/Vs. This value is comparable, but slightly higher, than what has been reported previously for Bi$_2$Se$_3$ surface states[26,27]; when inserted in the criterion for the occurrence of SdH oscillations, $\omega_c\tau = \mu B \geq 1$ (ref. 28), it gives $B \geq 4$ T, which agrees with experiments (oscillations due to top surface Dirac electrons start at approximately half this field –implying $\mu$ ~ 5000 cm$^2$/Vs- probably because contact with SiO$_2$ slightly reduces $\mu$ at the bottom surface).

**Landau levels for Dirac fermions.** We now proceed to discussing the indexing of the Landau levels responsible for the SdH. For massless Dirac fermions, when $n=(N+1/2)B/(h/e)$ the $N^{th}$ LL is completely filled and the Fermi energy is located in between the $N^{th}$ and the $(N+1)^{th}$ LLs (here $n$ is the density of surface Dirac fermions, $N$ the integer LL index, and $B/(h/e)$ the LL degeneracy[29]). A quantitative interpretation of the data measured in our samples, however, requires an assumption to be made: it is a priori unclear whether maxima or minima of the measured oscillations correspond to having completely filled LLs at the surface, because both the conductance of the Dirac fermions and the parallel conductance depend on magnetic field. Since the background conductance is 10x larger than the conductance of the surface states, its contribution to the oscillations is likely to dominate. We therefore assume that the dominant contribution to the magneto-resistance oscillations is due to the background conductance itself, which is modulated by the formation of surface LLs. Specifically, with the flake being only 10 nm thick, carriers responsible for the background conductance are strongly coupled to the surface states (this is indicated by the fact that the diffusion constant of these carriers and of the Dirac surface fermions is approximately the same, see the Supplementary Note 4). When the Fermi level at the surface is inside (in between) LLs, the density of states at the Fermi energy associated to the surface LLs has a maximum (minimum), so that carriers responsible for the background conductance have maximum (minimum) probability to scatter, and the background resistance is also maximum (minimum; in other words, the formation of surface LLs modulates the diffusion constant of the carriers responsible for the background conductance). On this basis, we index the LLs by taking resistance minima in the oscillations (as illustrated by the white lines in Fig. 3a) and plot the index $N$ as a function of $1/B$. For massless Dirac fermions $N = \frac{n h}{e} \frac{1}{B} - \frac{1}{2}$, i.e., the LL index $N$ scales linearly with 1/B, and extrapolates to $-1/2$ for $\frac{1}{B} \to 0$.

The analysis of the value of $N$ extrapolated from the actual data though a linear fit $N = a\frac{1}{B} + b$ is shown in Fig. 4, for all values of gate voltage for which the quality of the SdH oscillations is sufficient to perform the quantitative analysis ($V_g$ between -80 and -40 V for holes and between +15 and +50 V for electrons). For the different value of $V_g$, we find that the extrapolation $b$ indeed fluctuates close to –0.5, and the average of $b$ over the entire gate voltage range is $<b> = -0.4 \pm 0.1$, as expected for massless

Dirac fermions[29]. Despite requiring an assumption (i.e., that the oscillations are dominated by the conductance in parallel to that of the Dirac fermion states) this analysis is important because it shows that a simple physical scenario exists, that is based on what has been established about our system (presence of surface Dirac fermions and of a large parallel conductance) and that is capable of accounting for our observations quantitatively, in an internally consistent way.

**Gate-tuned supercurrent through surface states**. Measurements at $B = 0$ and $T = 30$ mK, with the Al/Ti electrodes in the superconducting state ($T_c \sim 1$ K), show that the ability to gate-control the device is also essential to establish that the Dirac fermions contribute to mediate the superconducting correlations induced by the contacts. In the measured differential resistance, a large suppression is observed when the bias is decreased below twice the superconducting gap (Fig. 5a), a distinctive signature of Andreev reflection indicating the high transparency of the contacts[31]. At zero bias, a supercurrent is present (Fig. 5b), with critical current $I_c \sim 200$ nA. It is apparent that the critical current depends on the gate voltage, and that it has a minimum in correspondence of the charge neutrality point extracted from the LL fan diagram. This finding shows the ambipolar character of the supercurrent, and makes clear directly from the experimental data that at least part of the supercurrent is carried by either Dirac electrons or holes at the bottom $Bi_2Se_3$ surface (and not only by the carriers in the bulk states). By enabling to separate the contributions of the surface and bulk channels to superconducting transport, gate control of supercurrent certainly provides an effective tool for the investigation of proximity effect in 3D topological insulators (whose complete understanding –e.g., the very small $R_nI_c$ product, $\sim 15$ µV $<< 2\Delta/e \sim 300$ µV - Fig. 5c- goes beyond the scope of this paper).

**Discussion**

Our experiments clearly show how the use of a gate electrode to tune the surface density of charge carriers is a very effective tool to identify and explore the contribution to transport of surface Dirac fermions in devices based on 3D topological insulators. The possibility to tune the carrier density electrostatically, for instance, enables the observation of Dirac electron and hole transport in a single device, making it possible to control the filling of their Landau levels. It also allows the identification of transport properties associated with the two opposite material surfaces, owing to their different electrostatic coupling to the gate electrode. Finally, it is crucial to establish that Dirac fermions are capable of mediating superconducting correlations induced by the contacts, a conclusion drawn directly from the experimental data that –in the presence of a large parallel contribution to transport- would be difficult to draw in any other way. It appears from our experiments that the quality of the BiSe-based 3D topological materials currently available would greatly benefit from considerable improvements, which should aim at decreasing the amount of states present in the bulk bandgap of the crystals and at increasing the carrier mobility. Nevertheless, even with the quality of the existing materials, performing new and interesting transport experiments, to investigate the physics of Dirac surface fermions is certainly possible. These experiments should address, among other, the detailed behaviour of the Dirac fermions as a function of the crystal thickness[32,33] –where ARPES experiments have shown an interesting dimensional crossover to occur[34]- and the nature of proximity effect[24,25], both of which currently remain phenomena virtually unexplored experimentally.

## Methods

**Growth of $Bi_2Se_3$ single crystals.** Single crystals of $Bi_2Se_3$ were grown by a floating zone (FZ) method. Bismuth and selenium were weighted in stoichiometric ratio and sealed under vacuum (~$10^{-3}$ mbar) in quartz ampoules. According to the Bi – Se phase diagram, $Bi_2Se_3$ has a congruent melting point at 705 ºC [35]. In order to synthesize the compound, the mixture of elements was heated up to 750 ºC, kept at this temperature for 12 h, then quenched to room temperature. To assure a better mixing of the elements, during the isothermal time in the liquid state, the ampoule was shaken and rotated. This procedure was repeated three times. Finally the ampoule was heated up to 820 ºC for 24 hours and slowly cooled down to room temperature. At the end of this procedure, the sample was already formed by large single domains of good quality, suggesting a high growth rate for $Bi_2Se_3$. The sample was then crushed into small pieces and sealed inside a quartz ampoule with 5 mm of inner diameter and a conical bottom, which was vertically mounted in a home-made 2-lamp furnace. The sample was moved downwards through the hot zone at a speed of ~0.33 mm/h, starting from the apex of the conical tip. In this way a single crystal started to grow in the tip. The ampoule was kept rotating all the time to avoid lateral thermal gradient. The advantage of the floating zone over other growth techniques consists in the large temperature gradient at the growth interface, the low traveling velocity, and the fact that the growing composition is prevented from local fluctuations in the melt composition. These differences can account for a different structural quality of the crystals –and in particular a lower density of defects- as compared to that of existing materials grown by different techniques. Large (1-10 mm size) crystals were cleaved from the solid rod formed inside the quartz. The high crystalline quality of the crystals was confirmed by X-ray diffraction that shows a single phase, single index diffraction pattern, and narrow rocking curves. The cell parameters were found to be in good agreement with those reported in literature.

**Device fabrication**. Thin layers of $Bi_2Se_3$ were obtained by exfoliation of single crystals using an adhesive tape, while monitoring the process under an optical microscope, similarly to what is done to extract graphene from graphite[36]. The layers were transferred onto a degenerately doped Si substrate (acting as a gate electrode) covered with a 300 nm layer of thermally grown $SiO_2$ (gate insulator), simply by pressing the adhesive tape onto the substrate. Most layers produced in this way appeared to be shiny when observed under the microscope, corresponding to a large thickness (several tens of nanometers or more). In a few cases, however, thinner layers (characterized by a darker colour) were also found (the thinnest layer that we observed had a thickness of approximately 4 nm). These thinner layers were selected for the fabrication of devices. Compared to graphene, producing very thin $Bi_2Se_3$ layers is considerably more difficult because $Bi_2Se_3$ is more brittle than graphite, and the material tends to easily break into very small pieces during the exfoliation process. Devices were fabricated by conventional nano-fabrication techniques: electron-beam lithography was used to generate the contact pattern, and electron-beam evaporation followed by lift-off was used to define the Al/Ti (75/5 nm) electrodes.

**Electrical measurements.** The devices were cooled down using either a Kelvinox 400 dilution refrigerator or a He3 insert, mounted into a cryostat equipped with a 15 T superconducting magnet, and measured in the range between 30 mK and 250 K. The dilution refrigerator contained copper powder filters at the mixing chamber stage, necessary to observe a finite supercurrent (all wires were also π-filtered at room temperature). The electrical characteristics were measured using a lock-in technique (magnetotransport), dc I-V curves (supercurrent measurements) or a combination of the two (bias dependent differential resistance measurements). The charge density in the devices was controlled by applying a gate voltage to the p-doped silicon substrate.

**Temperature dependence of the resistance.** As characterization of the $Bi_2Se_3$ crystals used in our experiments, we have measured the temperature dependence (from 250 to 4.2 K) of the resistance for two devices fabricated as described in Section 2, and compared the results to measurements reported in the literature. The two devices were fabricated using nearby layers (~100 μm apart) found on a same substrate, implying that they originate from a same region of the starting bulk $Bi_2Se_3$ crystal. The thickness of one of the layers was approximately 4 nm (as estimated using a Keyence$^{TM}$ microscope equipped with a laser interferometer) and that of the other was approximately 10 nm. The temperature dependence of the resistance is shown in Supplementary Figure S1 (unfortunately the 4 nm device broke early after starting the experiments). Although both layers were rather highly conducting ($\rho \sim 1$ m$\Omega$ cm, in the range reported in the literature[20-23]), the thinnest one was found to exhibit a weak insulating temperature dependence (resistance slightly increasing with lowering $T$), whereas in the thicker one the resistance decreased with lowering $T$. This difference may indicate the opening of a gap in the surface states due to hybridization of the Dirac bands on opposite surfaces, which for a 4 nm thickness should lead to a gap of about 10 meV (for a 10 nm thickness the gap is predicted to be negligible).

The large conductivity observed in nominally undoped $Bi_2Se_3$ is usually attributed to the presence of unintentional electron doping[21], which pins the Fermi level inside the bulk conduction band. Our gate dependent studies, which exhibit the charge neutrality point at only approximately -10 V, show that our devices are also slightly electron doped. Nevertheless, the gate voltage dependence of the resistance also indicates that, contrary to what is most frequently reported in the literature[20-23], in our thin layers the Fermi level at the surface is not located in the bulk conduction band, but inside the gap. Note that from the Fourier transform of the SdH oscillations we can conclude that the density of surface carriers at $V_g$=0 V is approximately the same on the two surfaces (see main text), implying that the Fermi level is inside the gap of the bulk bands also at the surface far away from the gate. Since the Thomas-Fermi screening length is not much smaller than the thickness of the crystal, we conclude that the Fermi level is inside the gap throughout the thickness of our devices. Note that, even though the observed behaviour is different from that reported from studies of transport through bulk crystals, it is in line with the observation reported by Steinberg et al. (see Ref. [26] who studied transport through gated exfoliated flakes; their devices are comparable but somewhat thicker than ours –the thinnest flake reported in Ref. [26] is approximately 16 nm thick).

Since in our devices the Fermi level is located in the bulk gap, the electrical conductance that we measure is mediated by both the surface Dirac fermions, and by carriers occupying states in the impurity band inside the gap, whose presence is established in the literature. The impurity band originates from defect states with a large Bohr radius –about 4-5 nm, owing to the large relative dielectric constant of the material; $\varepsilon$~110 [21]. This large radius causes a strong overlap of the impurity states, as well as a strong overlap of the impurity states with the surface Dirac fermion states (the Bohr radius of the impurity states is also comparable to the $Bi_2Se_3$ layer thickness), resulting in a high conductivity in the impurity band. The large value of $\varepsilon$ also suppresses Coulomb effects which are usually the dominant mechanism lowering the conductivity of impurity bands, and tend to cause a more insulating-like behaviour through the opening of a Coulomb gap.

Note, finally, that the density of bulk impurity band states that we have extracted from the data, $\nu_{Bulk}$ ~7 $10^{19}$ $cm^{-3}eV^{-1}$ (see main text), is comparable to the density of states of the impurity band that forms in degenerately doped silicon. For silicon, an impurity band (that causes the material to be conducting at very low temperature) develops at doping concentrations of a 3-4 $10^{19}$ $cm^{-3}$ [37]. The states responsible for this impurity band are close to either the valence or the conduction band (depending on the sign of the dopants), and they are distributed in a fraction of 1 eV (corresponding to the gap of silicon). The resulting density of states (equal to the concentration divided by this energy range) coincides with the value that we estimated for $Bi_2Se_3$. As mentioned above, the conductivity in the impurity band is much larger in $Bi_2Se_3$ as compared to silicon, because of the much larger dielectric constant which suppresses Coulomb gap effects.

# References

1. Kane, C. L. & Mele, E. J. Quantum Spin Hall Effect in Graphene. *Phys. Rev. Lett.* **95**, 226801 (2005).

2. Bernevig, B. A., Hughes, T. L. & Zhang, S.-C. Quantum Spin Hall Effect and Topological Phase Transition in HgTe Quantum Wells. *Science* **314,** 1757-1761 (2006).

3. König, M. *et al.* Quantum Spin Hall Insulator State in HgTe Quantum Wells. *Science* **318,** 766-770 (2007).

4. Roth, A. *et al.* Nonlocal Transport in the Quantum Spin Hall State. *Science* **325,** 294-297 (2009).

5. Fu, L., Kane, C. L. & Mele, E. J. Topological Insulators in Three Dimensions. *Phys. Rev. Lett.* **98,** 106803 (2007).

6. Fu, L., and Kane, C. L. Topological insulators with inversion symmetry. *Phys. Rev. B* **76,** 045302 (2007).

7. Moore, J. E. & Balents, L. Topological invariants of time-reversal-invariant band structures. *Phys. Rev. B* **75,** 121306 (2007).

8. Qi, X.-L., Hughes, T. L. & Zhang, S.-C. Topological field theory of time reversal invariant insulators. *Phys. Rev. B* **78,** 195424 (2008).

9. Zhang, H. J. *et al.* Topological insulators in $Bi_2Se_3$, $Bi_2Te_3$, and $Sb_2Te_3$ with a single Dirac cone on the surface. *Nature Physics* **5,** 438-442 (2009).

10. Hsieh D. *et al.* A topological Dirac insulator in a quantum spin Hall phase. *Nature* **452,** 970-974 (2008).

11. Hsieh, D. *et al.* A tunable topological insulator in the spin helical Dirac transport regime. *Nature* **460**, 1101-1105 (2009).

12. Xia, Y. *et al.* Observation of a large-gap topological-insulator class with a single Dirac cone on the surface. *Nature Physics* **5,** 398-402 (2009).

13. Cheng P. *et al.* Landau Quantization of Topological Surface States in $Bi_2Se_3$. *Phys. Rev. Lett.* **105,** 076801 (2010).

14. Hanaguri, T., Igarashi, K., Kawamura, M., Takagi, H., Sasagawa, T. Momentum-resolved Landau-level spectroscopy of Dirac surface state in $Bi_2Se_3$. *Phys. Rev. B* **82,** 081305 (2010).

15. Qu, D.-X., Hor, Y. S., Xiong, J., Cava R. J. & Ong, N. P. Quantum Oscillations and Hall Anomaly of Surface States in the Topological Insulator $Bi_2Te_3$. *Science* **329,** 821-824 (2010).

16. Analytis, J. G. *et al.* Two-dimensional surface state in the quantum limit of a topological insulator. *Nature Physics* **6,** 960-964 (2010).

17. Brüne, C. *et al.* Quantum Hall Effect from the Topological Surface States of Strained Bulk HgTe. *Phys. Rev. Lett.* **106,** 126803 (2011).

18. Ren, Z., Taskin, A. A., Sasaki, S., Segawa, K., Ando, Y. Large bulk resistivity and surface quantum oscillations in the topological insulator $Bi_2Te_2Se$. *Phys. Rev. B* **82,** 241306 (2010).

19. Checkelsky, J. G., Hor, Y. S., Cava, R. J., Ong, N. P. Bulk Band Gap and Surface State Conduction Observed in Voltage-Tuned Crystals of the Topological Insulator $Bi_2Se_3$. *Phys. Rev. Lett.* **106,** 196801 (2011).

20. Checkelsky, J. G. *et al.* Quantum Interference in Macroscopic Crystals of Nonmetallic $Bi_2Se_3$. *Phys. Rev. Lett.* **103,** 246601 (2009).

21. Analytis, J. G. *et al.* Bulk Fermi surface coexistence with Dirac surface state in $Bi_2Se_3$: a comparison of photoemission and Shubnikov-de Haas measurements. *Phys. Rev. B* **81,** 205407 (2010).

22. Eto, K., Ren, Z., Taskin, A. A., Segawa, K. & Ando, Y. Angular dependent oscillations of the magnetoresistance in $Bi_2Se_3$, due to the three-dimensional bulk Fermi surface. *Phys. Rev. B* **81**, 195309 (2010).

23. Butch, N. P. *et al.* Strong surface scattering in ultrahigh-mobility $Bi_2Se_3$ topological insulator crystals. *Phys. Rev. B* **81,** 241301 (2010).

24. Fu, L. & Kane, C. L. Probing Neutral Majorana Fermion Edge Modes with Charge Transport. *Phys. Rev. Lett.* **102**, 216403 (2009).

25. Akhmerov, A. R., Nilsson, J. & Beenakker, C. W. J. Electrically Detected Interferometry of Majorana Fermions in a Topological Insulator. *Phys. Rev. Lett.* **102**, 216404 (2009).

26. Steinberg, H., Gardner, D. R., Lee, Y. S. & Jarillo-Herrero, P. Ambipolar Electric Field Effect in Metallic $Bi_2Se_3$. *Nano Lett.* **10**, 5032-5036 (2010).

27. Cho, S., Butch, N., Paglione, J., Fuhrer, M. Insulating Behavior in Ultrathin Bismuth Selenide Field Effect Transistors. *Nano Lett.* **11,** 1925-1927 (2011).

28. Ashcroft, N. W. & Mermin, N. D. *Solid State Physics*. Saunders College Publishing (1976).

29. Castro Neto, A. H., Guinea, F., Peres, N. M. R., Novoselov, K. S. & Geim, A. K. The electronic properties of graphene. *Rev. Mod. Phys*. **81,** 109-162 (2009).

30. Alpichshev, Z. *et al*. STM imaging of impurity resonances on $Bi_2Se_3$. *arXiv*:1108.0022 (2011).

31. Blonder, G. E., Tinkham, M. & Klapwijk, T. M. Transition from metallic to tunneling regimes in superconducting microconstrictions: Excess current, charge imbalance, and supercurrent conversion. *Phys. Rev. B* **25**, 4515-4532 (1982).

32. Liu, C.-X. *et al.* Oscillatory crossover from two-dimensional to three-dimensional topological insulators. *Phys. Rev. B* **81**, 041307 (2010).

33. Shan, W.-Y., Lu, H.-Z. & Shen, S.-Q. Effective continuous model for surface states and thin films of three-dimensional topological insulators. *N. J. Phys*. **12,** 043048 (2010).

34. Zhang, Y. *et al.* Crossover of the three-dimensional topological insulator $Bi_2Se_3$ to the two-dimensional limit. *Nature Physics* **6,** 584-588 (2010).

35. Offergeld, G, and Van Cakenberghe, J. Stoichiometry of Bismuth Telluride and related compounds. *Nature* **184**, 185 (1959).

36. Novoselov, K. S. *et al.* Electric field effect in atomically thin carbon films. *Science* **306**, 666 (2004).

37. Sze, S. and Kwok, Ng. Physics of Semiconductor Devices. 3rd Ed. (Wiley, New Jersey, 2007).

**Acknowledgments** We gratefully acknowledge A. Ferreira for technical support, I. Martin and M. Büttiker for continued discussions on the subject of topological insulators, and D. van der Marel for encouragement and support. This work is partially financially supported by the Swiss Center of Excellence MaNEP and by the Swiss National Science Foundation (project 200021_121569).

**Author Contributions** J.B.O. fabricated the devices for which N.J.G.C. set up the electron-beam lithography system. B.S. set-up the cryogenic systems used to perform the low temperature experiments. B.S. and J.B.O. performed all the measurements. B.S., J.B.O., J.L., and A.F.M. analysed and interpreted the data. A.U. and E.G. grew the $Bi_2Se_3$ crystal. A.F.M. planned the research, supervised the work, and wrote the paper. All authors discussed the results.

**Competing Interests**  The authors declare that they have no competing financial interests.

**Figure Captions**

**Figure 1. Electrostatically gated devices based on a 3D topological insulator.** (a) Optical microscope image of a thin layer of $Bi_2Se_3$ on a $Si/SiO_2$ substrate, produced by exfoliation of a single crystal. The thickness is estimated to be ~ 10 nm, from measurements done with a Keyence[TM] microscope equipped with a laser interferometer. (b) Optical microscope image of the same layer with three Al/Ti (75/5 nm) contacts, forming two superconducting junctions. The contact separation L is approximately 400 nm; the width W of the two junctions is approximately 3.2 and 2 µm, respectively (all data shown in this paper have been obtained from the junction formed between the central and the left electrode; the other junction shows an essentially identical behavior). For both images the scale bar is 2 $\mu$m. (c) Schematic representation of the complete device, with the degenerately doped silicon wafer acting as a gate (the $SiO_2$ thickness is approximately 300 nm). A family of Dirac fermions resides on each surface of the $Bi_2Se_3$ layers, represented by the so-called Dirac cones.

**Figure 2. Gate voltage and magnetic field dependence of the normal state resistance.** (a) Gate voltage dependence of the normal state resistance. The maximum at $V_g$~-10 V, provides a first indication of ambipolar transport. The $V_g$ dependence originates from the modulation of the conductivity of the surface close to the gate, and can be quantitatively described as $\delta\sigma(V_g) = n_{Dirac}(V_g)e\mu$ ($n_{Dirac}(V_g)$ is given by Eq. 1) with $\mu$=2500 cm$^2$/Vs (inset of panel (a): the yellow line corresponds to the calculated $\delta\sigma(V_g)$). (b) Full dependence of the resistance on $V_g$ and magnetic field B (the measurements are performed by sweeping $V_g$ at fixed $B$, for different values of $B$). As a function of $B$ the resistance increases approximately linearly, exhibiting clear oscillations that are enhanced by deriving the data with respect to either $V_g$ or $B$. (c) Oscillations in $-\dfrac{d^2R}{dB^2}$ ($V_g$=-62 V), periodic in $1/B$ as it is typical for the Shubnikov-de Haas effect (for small oscillations superimposed on a slowly varying background $-\dfrac{d^2R}{dB^2}$ has the same phase as the oscillations in the resistance itself). All data shown in this figure were measured at 4.2 K.

**Figure 3. Fan diagram of Landau levels of Dirac electrons and holes.** (a) Plot of $-\dfrac{d^2R}{dB^2}$ (in arbitrary units) as a function of $B$ and $V_g$ (fan diagram). The features dispersing with $V_g$ originate from the formation of Landau levels of Dirac fermions on the bottom surface of the Bi$_2$Se$_3$ crystal. The features with a $V_g$-independent position are associated to Landau levels on the top surface. The white dashed lines (and the number indexing the LLs) indicate the condition for the Fermi level to fall between two Landau levels for holes on the bottom surface (see discussion in the main text). They correspond to $B = n\dfrac{h}{e}\dfrac{1}{N+\frac{1}{2}}$ ($n$ is the density of Dirac holes), as expected for Dirac fermions. A similar indexing also works for electrons (i.e., positive gate voltage). (b) $-\dfrac{d^2R}{dV_g^2}$ plot showing dispersing features corresponding to those observed in the $-\dfrac{d^2R}{dB^2}$ plot (the $V_g$-independent features disappear when taking the $V_g$ derivative). An asymmetry between electron and holes at high magnetic field –already visible in the $-\dfrac{d^2R}{dB^2}$ plot- is present, which is likely to originate from the opening of a small Zeeman

gap[16], which shifts the position of the zero-energy LL to either positive or negative energies depending on the sign of the *g*-factor. (c) Fourier spectrum of $-\frac{d^2 R}{dB^2}$, as a function of $V_g$ and frequency $B_F$. From the position maxima in the spectrum we extract the density of carriers according to the relation $n = B_F e/h$. Again, the features dispersing with $V_g$ originate from Landau levels of Dirac fermions on the bottom surface, and the $V_g$-independent features (pointed to by the red arrow), from states on the top surface. At $V_g=0$ V, the peaks of the dispersing and non-dispersing features occur approximately at the same frequency, implying that the density of electrons on the two surfaces of the crystal are also the same. The blue line is the frequency of the SdH oscillations calculated using Eq. (1) ($B_F = n_{Dirac}(V_g)h/e$) with no free parameter, in good agreement with the gate dependent position of the (broad) peak in the Fourier spectrum.

**Figure 4. Massless Dirac character of surface fermions.** Main panel: Landau level index *N* versus *1/B* plotted for selected values of gate voltage, corresponding to different concentrations of holes (red symbols) and electrons (blue symbols). According to the relation characteristic for massless Dirac fermions, $N = \frac{n h}{e} \frac{1}{B} - \frac{1}{2}$, where *n* is the density of charge carriers (electrons or holes), whereas for normal fermions $N = \frac{n h}{e} \frac{1}{B}$. Therefore, looking at how *N* extrapolates for $\frac{1}{B} \to 0$ allows us to discriminate between massless Dirac and normal fermions. The continuous lines are linear least square fits to the data ($N = a\frac{1}{B} + b$). They show that as $\frac{1}{B} \to 0$, *N* extrapolates to close to *-1/2* as expected for massless Dirac fermions. The inset shows the extrapolation value (*b*) obtained by least square fitting for approximately 20 different values of gate voltage (in the region with Vg in between ~ -40 and +15 V, the oscillations are not sufficiently pronounced to perform the analysis). The data show that $b$ fluctuates close to -1/2 (and not around 0). Consistently, the average of $b$ over gate voltage is $<b> = -0.4 \pm 0.1$.

**Figure 5. Gate-tunable supercurrent.** (a) Differential resistance of the junction in the superconducting state ($B$=0, $T$=30 mK), as a function of bias voltage $V$ ($V_g$=+50 V). A large decrease is seen upon lowering $V$ below $2\Delta/e$ ~300 µeV, due to Andreev reflection (the sharp feature around $V$=0 originates from the presence of the supercurrent; the origin of the split peak at $V$~$2\Delta/e$ is currently unknown). (b) Junction *I-V* curves showing a gate-tunable supercurrent. The critical current exhibits a minimum at $V_g$~-10 V, corresponding to the $V_g$ value at which the maximum of the normal state resistance occurs. (c) $R_n I_c$ product of the junction obtained using for $R_n$ the measured normal state resistance. Its value (~15 µeV) is about 20 times smaller than expected ($R_n I_c$ ~$2\Delta/e$~300 µeV).

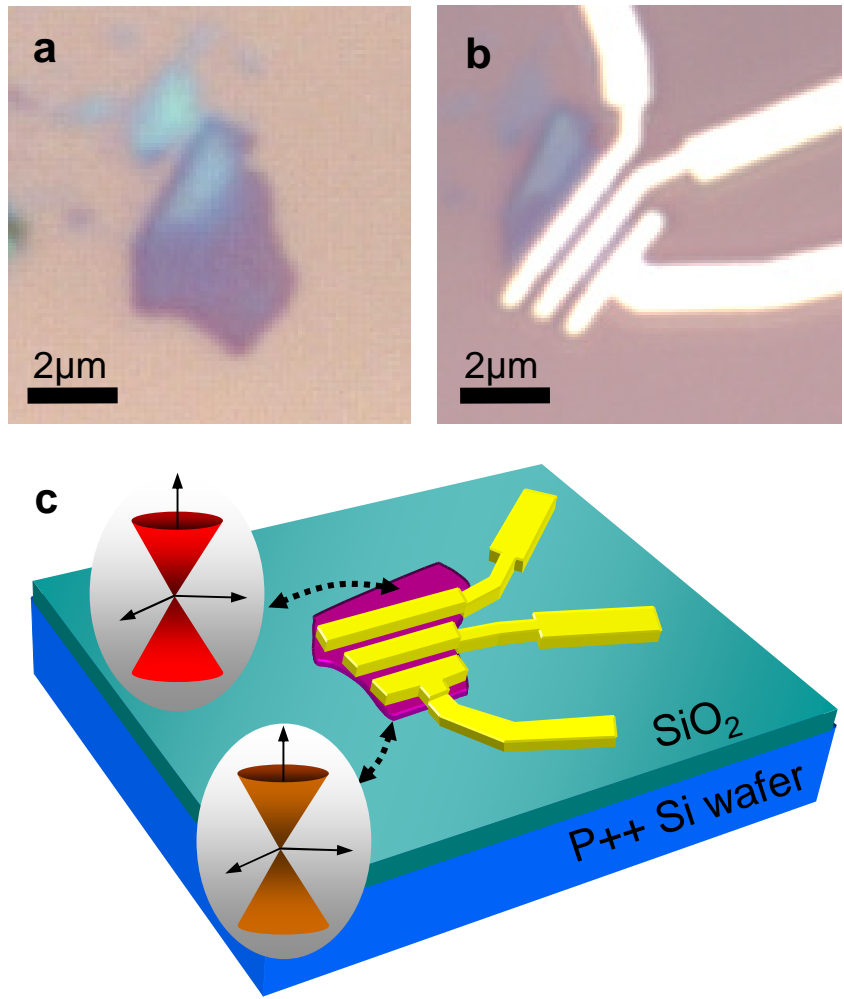

Figure 1

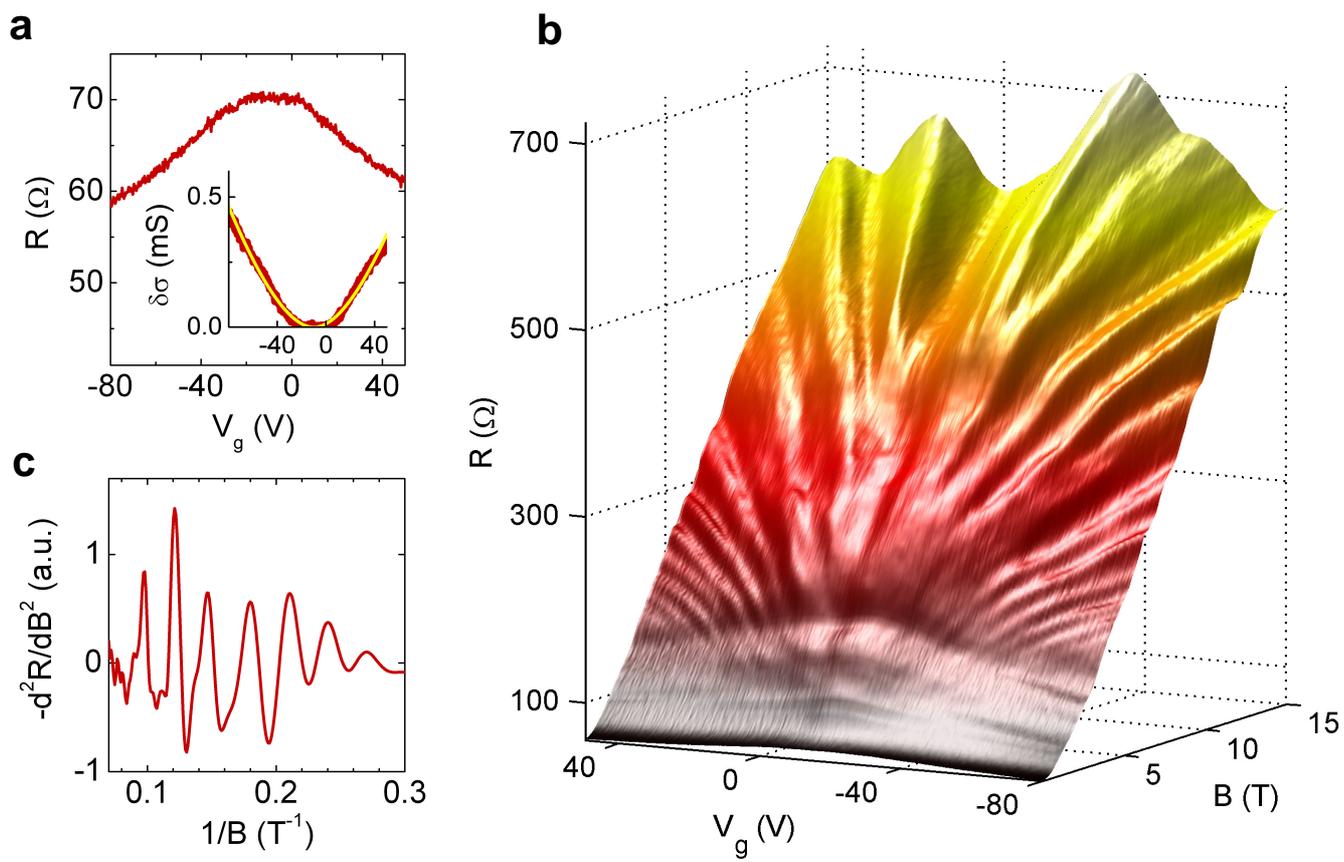

Figure 2

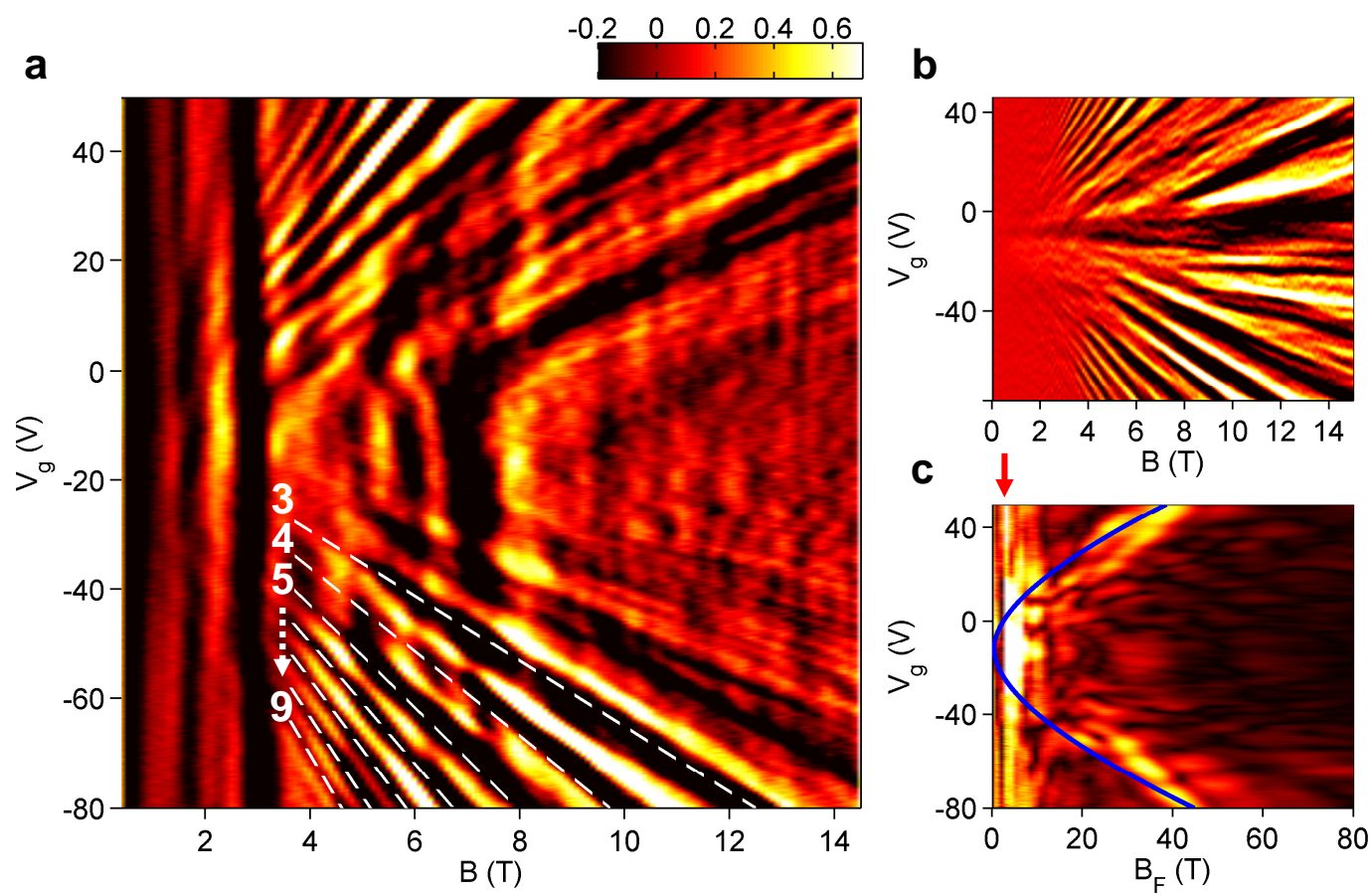

Figure 3

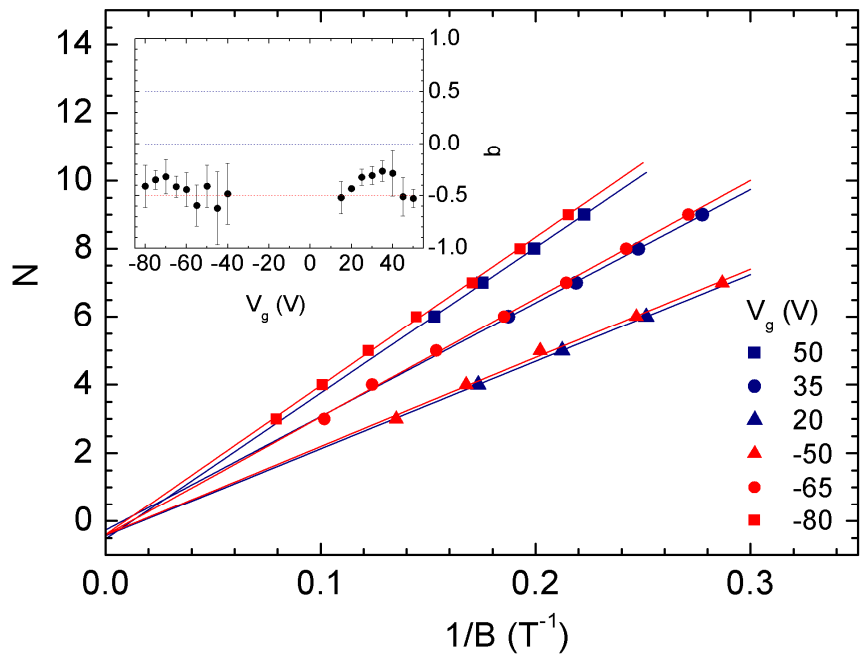

Figure 4

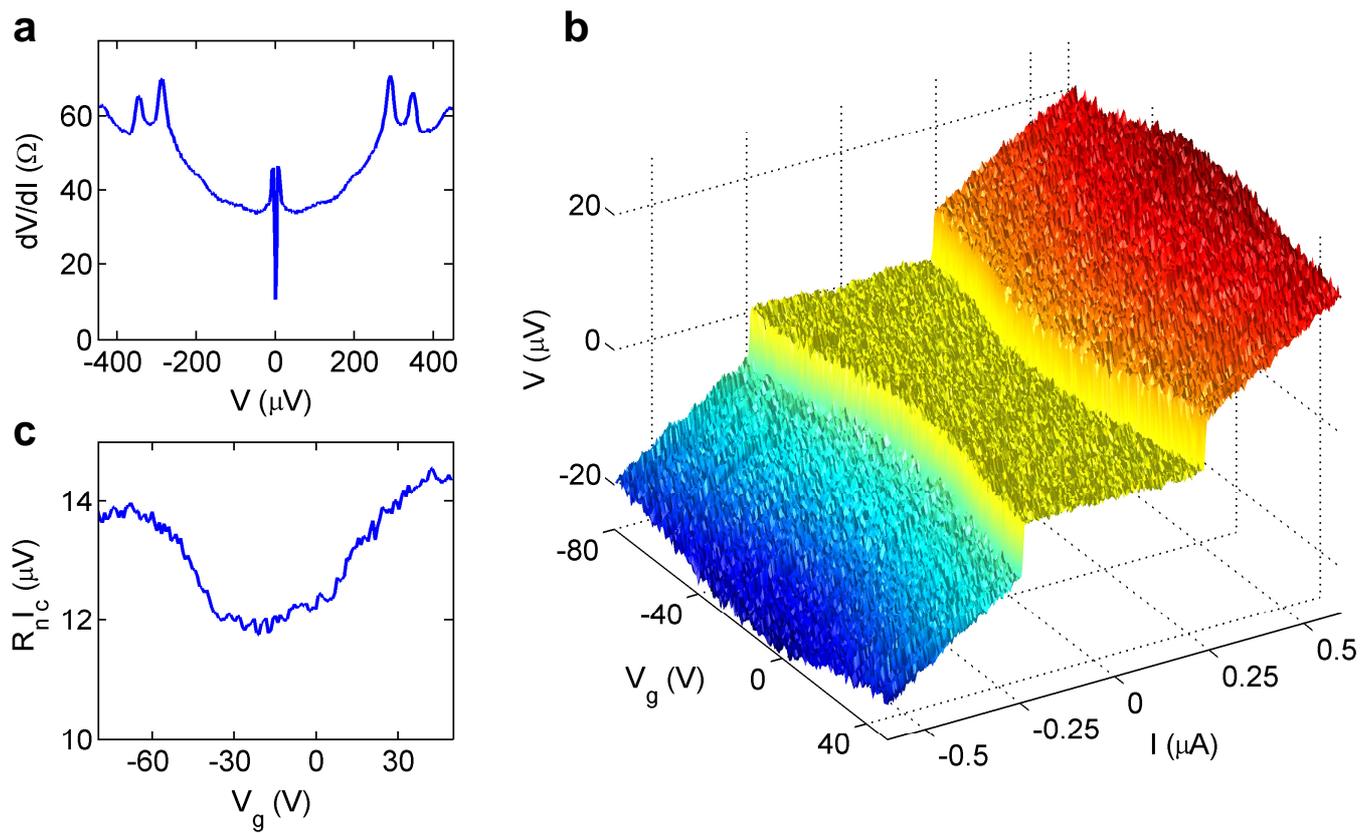

Figure 5